\documentclass[aps, prb, superscriptaddress, twocolumn, a4paper,10pt]{revtex4-2}
\usepackage{hyperref}
\usepackage{graphicx}
\usepackage{amsmath}
\usepackage{amssymb}
\usepackage{amsfonts}   
\usepackage{braket}
\usepackage{color}

\usepackage{dsfont}

\renewcommand{\Im}{\operatorname{Im}}
\newcommand{\Tr}{\mathrm{Tr}}

\begin{document}
\title{Localization and spectrum of quasiparticles in a disordered fermionic Dicke model}

\author{Sebastian Stumper}
\affiliation{Institute of Physics, University of Freiburg, Hermann-Herder-Str. 3, 79104 Freiburg, Germany}

\author{Junichi Okamoto}
\affiliation{Institute of Physics, University of Freiburg, Hermann-Herder-Str. 3, 79104 Freiburg, Germany}
\affiliation{EUCOR Centre for Quantum Science and Quantum Computing, University of Freiburg, Hermann-Herder-Str. 3, 79104 Freiburg, Germany}

\date{\today}

\begin{abstract}
	We study a fermionic two-band model with the interband transition resonantly coupled to a cavity. This model was recently proposed to explain cavity-enhanced charge transport, but a thorough characterization of the closed system, in particular localization of various excitations, is lacking. In this work, using exact diagonalization, we characterize the system by its spectrum under various filling factors and variable disorder. As in the Dicke model, the effective light-matter coupling scales with the square root of the system size. However, there is an additional factor that decreases with increasing doping density. The transition from the weak-coupling regime to the strong-coupling regime occurs when the effective light-matter coupling is larger than the electronic bandwidth. Here, the formation of exciton-polaritons is accompanied by the formation of bound excitons. Photon spectral functions exhibit significant weights on the in-gap states between the polaritons, even without disorder. Finally, while the localization of electron-hole excitations in a disordered system is lifted by strong coupling, the same is not true for free charges, which remain localized at strong and even ultrastrong coupling. Based on this finding, we discuss scenarios for charge transport.
\end{abstract}

\maketitle
\section{Introduction}\label{sec:introduction}

Strong and ultrastrong coupling between light and matter has opened up new avenues in chemistry and physics~\cite{friskkockumUltrastrongCouplingLight2019,hertzogStrongLightMatter2019,garcia-vidalManipulatingMatterStrong2021,reitzCooperativeQuantumPhenomena2022}. In these regimes, various fundamental matter excitations can hybridize with photons to form so-called polaritons, and hence the basic properties of a physical system can be altered. For example, polaron-photon coupling can modify chemical reaction paths and enhance or reduce the reaction rate~\cite{herreraCavityControlledChemistryMolecular2016,galegoSuppressingPhotochemicalReactions2016,herreraMolecularPolaritonsControlling2020,nagarajanChemistryVibrationalStrong2021}. Bose-Einstein condensation and lasing is observed in a system of exciton-polaritons~\cite{dengPolaritonLasingVs2003,kasprzakBoseEinsteinCondensation2006,keelingCollectiveCoherencePlanar2007,byrnesExcitonPolaritonCondensates2014,fraserPhysicsApplicationsExciton2016}. Further possible applications include quantum information processing~\cite{sanvittoRoadPolaritonicDevices2016,kavokinPolaritonCondensatesClassical2022} and material engineering~\cite{ciutiQuantumVacuumProperties2005,laussyExcitonPolaritonMediatedSuperconductivity2010,kolaricStrongLightMatter2018,bartoloVacuumdressedCavityMagnetotransport2018a,naudet-baulieuDarkVerticalConductance2019,quachSuperabsorptionOrganicMicrocavity2022}. The physical platform is not limited to an optical cavity. Similar effects can be achieved in circuit~\cite{blaisCircuitQuantumElectrodynamics2021} or plasmonic~\cite{tameQuantumPlasmonics2013,tormaStrongCouplingSurface2014,ginzburgCavityQuantumElectrodynamics2016,bozhevolnyiCaseQuantumPlasmonics2017} quantum electrodynamics.

Among the various applications is the manipulation of transport processes in organic materials. For example, strong light-matter coupling can enhance energy transfer processes~\cite{zhongNonRadiativeEnergyTransfer2016,zhongEnergyTransferSpatially2017a,garcia-vidalLongdistanceOperatorEnergy2017,reitzEnergyTransferCorrelations2018}. The underlying mechanism is by now well understood in terms of a cavity-mediated long-range hopping of excitons~\cite{feistExtraordinaryExcitonConductance2015,schachenmayerCavityEnhancedTransportExcitons2015,chavezDisorderEnhancedDisorderIndependentTransport2021,sommerMolecularPolaritonicsDense2021,engelhardtUnusualDynamicalProperties2022}. Important steps towards manipulating charge transport were taken by experiments in the Ebbesen group~\cite{orgiuConductivityOrganicSemiconductors2015,nagarajanConductivityPhotoconductivityPType2020}. Here, the strong coupling between a surface plasmon and organic semiconductors (both $n$-type and $p$-type) enhances the electric conductivity. Since electrons in organic semiconductors, in general, suffer from large static and dynamic disorder, their mobility is ascribed to incoherent diffusive transport~\cite{coropceanuChargeTransportOrganic2007,basslerChargeTransportOrganic2011,ciuchiTransientLocalizationCrystalline2011,liGeneralChargeTransport2021}. The enhancement of electric conductivity indicates a novel type of transport mechanism, because effective long-range hopping, as in the case of excitons, is impossible~\cite{zebIncoherentChargeTransport2022}. Thus, it is an important problem to understand the microscopic processes.

By now, a few theoretical studies have manifested the enhanced electronic transport via strong light-matter coupling. In Refs.~\cite{hagenmullerCavityEnhancedTransportCharge2017,hagenmullerCavityassistedMesoscopicTransport2018}, a mechanism is described where the coupling opens up an additional transport channel through another band. At the same time, the role of source and drain terminals are emphasized. The original publication on the experiment maintains that the electronic wavefunctions are delocalized as a result of the strong hybridization of excitons and plasmons, which presumably enhances the electronic conductivity~\cite{orgiuConductivityOrganicSemiconductors2015}. However, the mechanism behind this delocalization remained an open question. Further studies have considered, the charge transfer rates from donors to acceptors~\cite{wellnitzQuantumOpticsApproach2021} or between molecules of the same type~\cite{zebIncoherentChargeTransport2022}.

In this paper, we study disordered two-band electrons strongly coupled to a cavity mode in order to elucidate the relationship among the excitation spectrum, localization and the mobility. By exact diagonalization, we first calculate the energy spectrum of the model and clarify the effects of the rotating-wave approximation (RWA), electron filling, and disorder. We show that, while the spectrum at half-filling depends on the light-matter coupling $g$ through $\tilde{g}=g \sqrt{L}$ as in the Dicke model, the doping modifies the scaling form.

There are various regimes depending on the strength of the light-matter coupling. At weak coupling $\tilde{g}$, few eigenstates have a significant photon content. As $\tilde{g}$ increases, at an intermediate point $\tilde{g}^\text{macro}$, the cavity excitation is shared by a macroscopic number of states. Shortly after this point, cavity-mediated electron-hole attraction leads to the formation of excitons. The threshold for the strong coupling regime, $\tilde{g}_\text{th}^\text{pol}$, is then marked by the formation of well-defined exciton-polariton states; it occurs when the effective coupling $\tilde{g}$ is larger than the bandwidth $W$. All of these effects are captured by the RWA. In the ultrastrong coupling regime, $\tilde{g} \gtrsim\tilde{g}_\text{th}^\text{US}$, where the RWA breaks down, the mixing of states with different photon numbers becomes important. Deep inside the ultrastrong coupling regime, there exists a transition to a superradiant phase. For realistic parameters for organic semiconductors, these energy scales follow 
\begin{equation}
	\tilde{g}^\text{macro} < \tilde{g}_\text{th}^\text{pol} \ll \tilde{g}_\text{th}^\text{US} \ll \tilde{g}_\text{c}^\text{SR}.
\end{equation}
We shall mainly focus on the strong and ultrastrong coupling regimes sufficiently below $\tilde{g}_\text{c}^\text{SR}$, since the existence of the superradiant phase may be prevented by terms that are not included in the model~\cite{bialynicki-birulaNogoTheoremConcerning1979a,natafNogoTheoremSuperradiant2010,debernardisCavityQuantumElectrodynamics2018,pilarThermodynamicsUltrastronglyCoupled2020}.

We further investigate the generalized inverse participation ratio, which quantifies the localization of different kinds of quasiparticles. It is shown that the localized charges, i.e., electrons and holes, remain localized under coupling to a cavity mode, and their transport behavior is marginally affected. In contrast, electron-hole excitations localized by disorder can be delocalized due to the light-matter coupling, as also shown in previous studies~\cite{schachenmayerCavityEnhancedTransportExcitons2015,feistExtraordinaryExcitonConductance2015}. Thus, we consider that the enhancement of electric conductivity is achieved through exciton-polariton formation and speculate that the dissociation of the exciton at the electrodes play an important role.

The rest of the paper is organized as follows. In Sec.~\ref{sec:model}, we present the model and explain the structure of the Hilbert space. The employed method is briefly discussed as well. Sec.~\ref{sec:energy} shows the energy spectrum of the model and compares various effects of the RWA, disorder, and doping. In Sec.~\ref{sec:LDOS}, we discuss the local density of states of various excitation and elaborate on the formation of excitons and exciton-polaritons. The localization properties of electrons and excitons under strong coupling are discussed in Sec.~\ref{sec:GIPR} by inverse participation ratios. Sec.~\ref{sec:conclusion} concludes the paper.

\section{Model}\label{sec:model}

\begin{figure}
	\centering
	\includegraphics[width=\columnwidth]{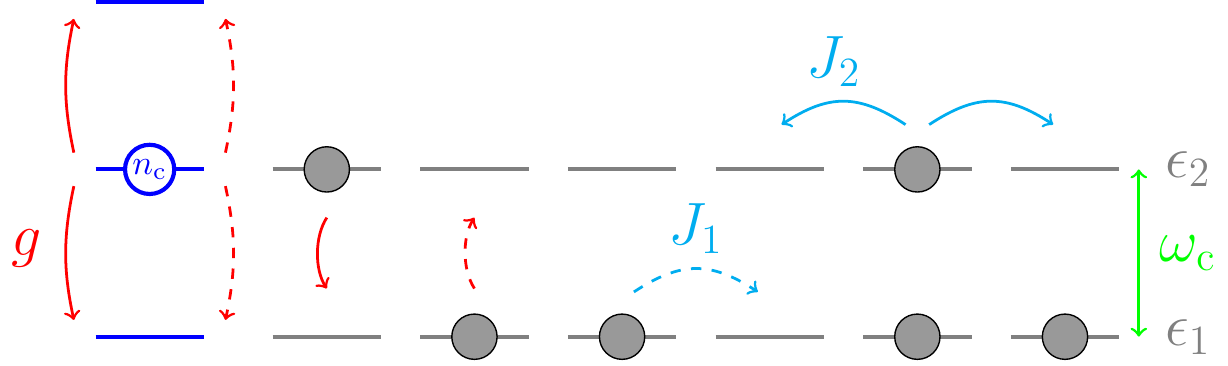}
	\caption{Schematic of the model indicating the cavity levels (left, blue) and the average local electronic levels (right, gray), where the cavity frequency $\omega_\mathrm{c}$ and the electronic energy gap $\epsilon_2-\epsilon_1$ are on resonance. The hopping strengths $J_{1,2}$ differ between both bands and the light-matter coupling $g$ can induce processes where electron-hole excitations are created (red dashed arrows) or annihilated (red solid arrows).}
	\label{fig:model}
\end{figure}

As a model of a one-dimensional organic semiconductor coupled to cavity photons or surface plasmons, we consider the following Hamiltonian
\begin{equation}
\begin{split}
H &= H_\text{el} + H_\text{cav} + H_\text{el-cav}, \label{eq:H_light_matter}
\end{split}
\end{equation}
consisting of a purely electronic part, a bosonic part, and an interaction term. These are, respectively, given by
\begin{subequations}
	\label{eq:H_terms}
	\begin{align}
	H_\text{el} &= - \sum_{\nu,r} J_{\nu,r} \left( c_{\nu,r}^\dagger c_{\nu,r+1} + c_{\nu,r+1}^\dagger c_{\nu,r} \right) \notag \\
	& \hspace{12pt} + \sum_{\nu,r} \epsilon_{\nu,r} n_{\nu,r} \label{eq:H_el} \\
	H_\text{cav} &= \omega_\text{c} a^\dagger a \label{eq:H_cav} \\
	H_\text{el-cav} &= g (a+a^\dagger) \sum_r \left[ c_{2,r}^\dagger c_{1,r} + c_{1,r}^\dagger c_{2,r} \right], \label{eq:H_el-cav}
	\end{align}
\end{subequations}
and schematically shown in Fig.~\ref{fig:model}. The labels represent lattice site ($r=1,\ldots,L$), lower and upper molecular orbital ($\nu=1,2$), and cavity ($\text{c}$). The hopping and on-site terms of the electrons can be site-dependent due to static disorder, i.e., $J_{\nu,r} = J_\nu + \delta J_{\nu,r}$ and $\epsilon_{\nu,r} = \epsilon_\nu + \delta \epsilon_{\nu,r}$. We fix the cavity frequency to be resonant with the average molecular excitation energy, $\omega_\text{c} = \epsilon_2 - \epsilon_1$. Note that the electrons are spinless, since at the fillings considered below not even on-site interaction effects between different species are important. 

The model represents a fermionic generalization of the Dicke model where each lattice site has four instead of two states, including doubly occupied and completely unoccupied molecules~\cite{hagenmullerCavityassistedMesoscopicTransport2018}. The Hilbert space $\mathcal{H}$ can, thus, be block-diagonalized according to the total electron number $0\leq N\leq 2L$, which is conserved by the Hamiltonian, i.e., $\mathcal{H}=\bigoplus_N \mathcal{H}_N$. The analogy to the Dicke model is most pronounced in the half-filled case $N=L$, where the electronic ground state at $g=0$ consists of the fully occupied lower orbitals and empty upper orbitals. In every other sector $\mathcal{H}_{N\neq L}$, any state involves doubly occupied or empty molecules. These molecules can be considered as carrying a conduction electron or a hole. In the present paper, the electron number is mainly restricted to $N=L-1,L,L+1$, i.e., the undoped, single-electron, or -hole states.

Each of these sectors can be characterized by the total number of excitations $M=n_\text{c}+n_\text{exc}$, where $n_\text{c}$ represents the number of photons and $n_\text{exc}$ is the number of excited electrons. For $N\leq L$, the latter is equal to the number of electrons in the upper orbitals, i.e., $n_\text{exc}=\sum_r \langle c^\dagger_{2,r} c_{2,r} \rangle \leq N$, while for $N\geq L$ it is given by the number of holes or unoccupied lower orbitals, i.e., $n_\text{exc}=\sum_r \langle c_{1,r} c^\dagger_{1,r} \rangle \leq 2L-N$.  

While the Hamiltonian does not conserve $M$, it does conserve the parity $M(\text{mod}\,2)$, which corresponds to the $\mathbb{Z}_2$ symmetry given by the transformation
\begin{equation}\label{eq:parity}
	a \leftrightarrow -a, \qquad c_{\nu,j} \leftrightarrow - c_{\nu,j}. 
\end{equation}
However, if the RWA is employed, the coupling Hamiltonian in Eq.~\ref{eq:H_el-cav} is simplified to
\begin{equation}
	H_\text{el-cav}^\text{RWA} = g \sum_r \left[ a c_{2,r}^\dagger c_{1,r} + a^\dagger c_{1,r}^\dagger c_{2,r} \right], \label{eq:H_el-cav_RWA}
\end{equation} 
and $M$ is conserved. The omitted terms are counter-rotating terms, which couple different M sectors.

Throughout the paper, the model parameters are chosen to be realistic for the experimentally relevant organic materials~\cite{orgiuConductivityOrganicSemiconductors2015}. In units of the average hopping $J_2$ in the upper band, we have $J_1=-J_2/2$, as an ad hoc choice reflecting the commonly asymmetric hopping of between upper orbitals compared to lower orbitals~\cite{coropceanuChargeTransportOrganic2007}. These, parameters determine the charge mobilities, along with disorder and dephasing terms. Furthermore, the dominating energy scale in organic semiconductors is the band gap, which is ten to a hundred times larger than the bandwidths. We chose the centers of the comparatively narrow bands to be on resonance with the cavity mode, i.e. $\epsilon_2-\epsilon_1=\omega_\text{c}=66 J_2$. In the transport experiments of Ref.~\cite{orgiuConductivityOrganicSemiconductors2015}, this gap was approximately equal to $2\text{eV}$, suppressing thermal excitation of charges.

For the disordered cases, we use normally distributed random variables $\delta J_{2, r}\sim\mathcal{N}(0.2 J_2)$, $\delta J_{1, r}\sim\mathcal{N}(0.5 J_2)$, $\delta \epsilon_{2, r}\sim\mathcal{N}(3 J_2)$, $\delta \epsilon_{2, r}\sim\mathcal{N}(5 J_2)$ as suggested in Ref.~\cite{orgiuConductivityOrganicSemiconductors2015}. Here, $\mathcal{N}(\sigma)$ is a normal distribution with variance $\sigma$. The bandwidth $W$ is thus mostly governed by disorder. Similarly to the choice of the hopping parameters, this corresponds to higher mobility in the upper band. Furthermore, it ensures a small localization length which facilitates the use of exact methods to study mobility in small systems.

\section{Energy eigenstates}\label{sec:energy}

\begin{figure*}[t!]
	\centering
	\includegraphics[width=\textwidth]{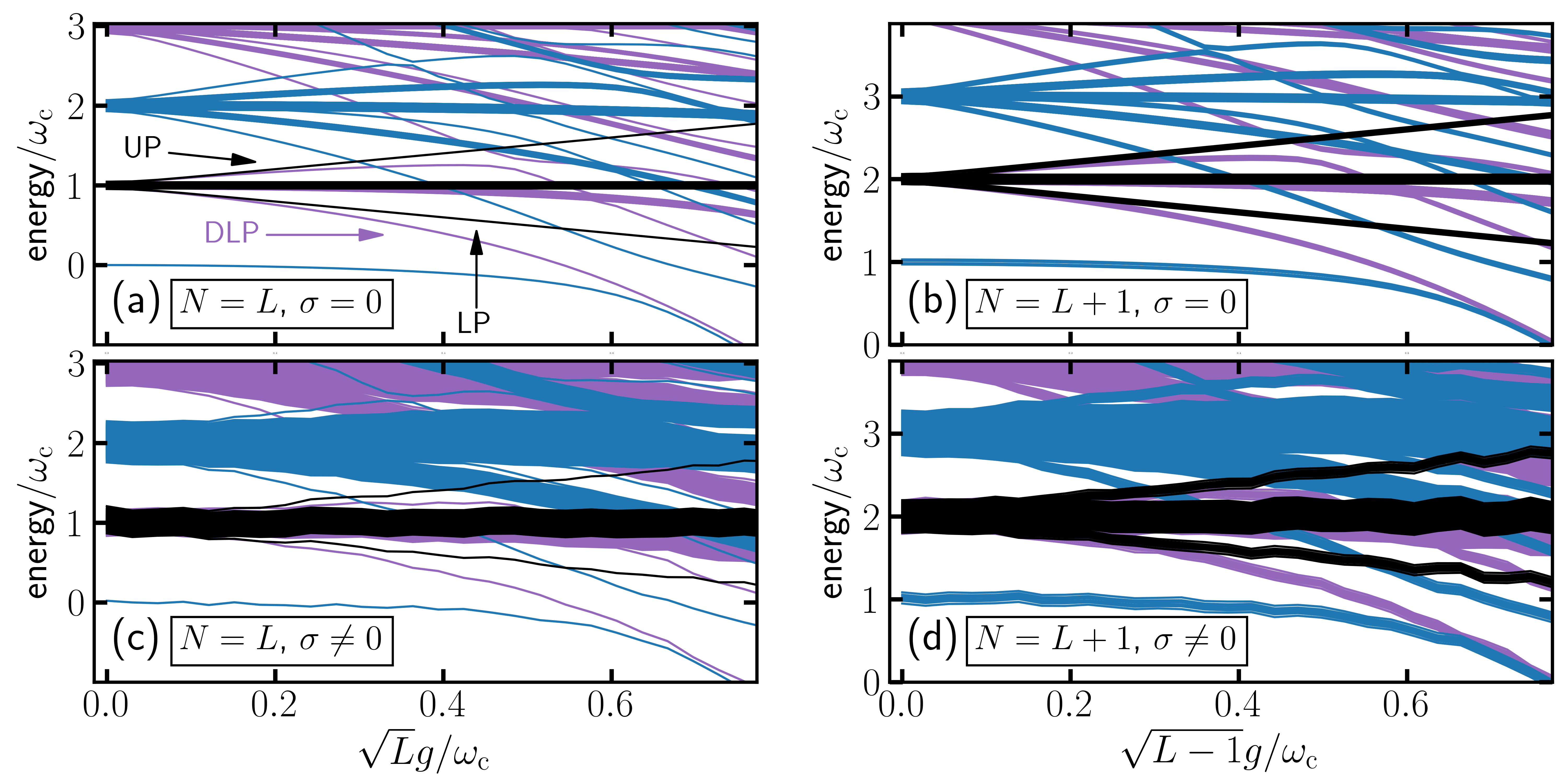}
	\caption{Energy spectrum of the model for $L=6$ computed for up to $M=8$ excitations, though only up to $M=4$ are shown. Half-filling $N=L$ [(a) and (c)] and a lightly doped case $N=L+1$ [(b) and (d)] are compared. The upper row [(a) and (b)] shows the homogeneous case, whereas the lower row [(c) and (d)] shows disorder-averaged spectra. Colors indicate sectors of even (blue) and odd (purple) parity. For comparison, the single excitation sector is shown with the RWA (black). Arrows in panel (a) indicate the lower/upper polariton (LP/UP) and the dressed lower polariton (DLP).}
	\label{fig:spec}
\end{figure*}

In this section we study the energy spectrum, and how it depends on filling, disorder and the presence of the counter-rotating terms. It shall be demonstrated that the spectrum of our Hamiltonian depends on the light-matter coupling $g$ and the system size $L$ only through the combination of
\begin{equation} \label{eq:effective_coupling}
	\tilde{g} \equiv g\sqrt{L(1-n_\text{d})}, 
\end{equation}	
where $n_\text{d}=|(N-L)/L|$ is the doping density, i.e., the density of conduction electrons or holes. This scaling form is similar to the one in the Dicke model, where the effective light-matter coupling is given by $g \sqrt{L}$, despite the difference of the local Hilbert spaces. In simple terms, the reason for this scaling factor is that only those molecular states with one electron, either in the upper or the lower orbital, couple to the cavity, while the doubly occupied and empty states do not. Accordingly, $L(1-n_\text{d})$ is the number of molecules that can couple to the cavity.

To further substantiate this scaling behavior, it is instructive to consider the single-excitation sector ($M=1$). In the case of zero disorder, electronic eigenstates are given by plane waves, $b^\dagger_{\nu,k} = L^{-1/2} \sum_r e^{i k r} c^\dagger_{\nu, r}$. This renders the coupling Hamiltonian as
\begin{equation}
	H_\text{el-cav} = g (a+a^\dagger) \left( \sum_k b_{2,k}^\dagger b_{1,k} + \mathrm{h.c.} \right).
\end{equation}
The mode created by
\begin{equation}\label{eq:bright_mode}
	B^\dagger = \sum_k b_{2,k}^\dagger b_{1,k} 
\end{equation} 
is commonly called the bright mode, as it couples to the cavity, while the modes that are orthogonal to it are dark modes. 

Let $\ket{\text{G}}=\ket{\text{FS}}\otimes\ket{0}_\text{c}$ denote the ground state at $g=0$ where $\ket{\text{FS}}=\prod_{(\nu,k)\in\text{FS}} b_{\nu,k}^\dagger \ket{0}_\text{el}$ is the Fermi sea with a given electron number $N$, and $\ket{0}_\text{el/c}$ represent the electronic and cavity vacua. The norm of the state $B^\dagger \ket{FS}$ 
is given by
\begin{equation}
	||B^\dagger \ket{FS}|| = \sqrt{L(1-n_\text{d})}.
\end{equation}
In order to express the light-matter coupling as a matrix element of $H_\text{el-cav}$ between orthonormal states in the single-excitation sector, one has to use
\begin{equation}\label{eq:states}
	\begin{split}
		\ket{B;0} &= \frac{1}{\sqrt{L(1-n_\text{d})}} B^\dagger \otimes \mathds{1} \ket{G}, \\
		\ket{0;1} &= \mathds{1} \otimes a^\dagger \ket{G}
	\end{split}
\end{equation}
as states with a single matter excitation or a single cavity excitation. Thus, the matrix element attains the anticipated scaling factor from Eq.~\eqref{eq:effective_coupling}
\begin{equation}\label{eq:rescaled_coupling}
	\braket{ 0;1 | H_\text{el-cav} | B;0 } = g \sqrt{L(1-n_\text{d})}.
\end{equation}

When the effective coupling $\tilde{g}$ is significantly larger than the electronic bandwidths $W$, the eigenstates of the single excitation sector can be well approximated in terms of upper and lower polaritons (UP and LP)
\begin{equation}\label{eq:polariton}
	\ket{P^\pm} \sim \ket{B;0} \pm \ket{0;1},
\end{equation}
as well as dark states. Counter-rotating terms have the effect of dressing these states with higher odd excitation numbers (DUP and DLP, where D stands for ``dressed''). In our model, the collective excitation can be considered as an exciton-polariton; electrons and holes from these excitons via a cavity-induced attractive interaction as shown below.

A second analytical approach that yields a scaling behavior as in Eq.~\eqref{eq:effective_coupling} is the mean-filed approximation of the cavity field. Within this approximation, there is a superradiant transition that breaks the parity symmetry of Eq.~\eqref{eq:parity} at a critical coupling 
\begin{equation}
	g_\text{c}^\text{MF,SR} \sim \frac{1}{\sqrt{L(1-n_\text{d})}}.
\end{equation}
For the calculation, we have taken the zero-temperature limit, followed by the flat band limit. Details are given in Appendix~\ref{app:MF}.

\subsection{Full energy spectrum}\label{sec:Spectrum}

In order to understand the effect of light-matter coupling on the energy levels, we compute the full spectrum for a small system size $L=6$ by exact diagonalization, such that the low energy part is converged with respect to the number of excitations $M$, even when $g$ is large. Here, we find that it is sufficient to include $M=8$ excitations to capture the corrections to the lowest energy states induced by the counter-rotating terms for $\tilde{g}\leq 0.8 \omega_\text{c}$. 

Energy levels corresponding to the lowest excitation numbers, $M\leq 4$, are shown with respect to appropriately rescaled coupling strengths in Fig.~\ref{fig:spec}. The immediate and most important conclusion from these spectra is that there are barely qualitative differences between half-filling $N=L$ [Fig.~\ref{fig:spec}(a)~and~(c) for the homogeneous and the disordered model] and a ``doped’’ state with one conduction electron, i.e. $N=L+1$ [Fig.~\ref{fig:spec}(b)~and~(d)].

All cases have in common that there are three regimes determined by the effective coupling $\tilde{g}$. First, focussing on the single excitation sector ($M=1$), the polariton states appear when $\tilde{g}$ is larger than a threshold that is determined by the bandwidth, 
\begin{equation}
	\tilde{g} > \tilde{g}_\text{th}^\text{pol}\gtrsim W,	
\end{equation}
marking the regime of strong coupling. However, the details of this transition are not fully clear from the spectrum and will be more investigated more carefully in Sec.~\ref{sec:LDOS_exc}. Next, for
\begin{equation}
	\tilde{g}\gtrsim \tilde{g}_\text{th}^\text{US} \approx 0.2\omega_\text{c}, 
\end{equation}
the exact energies show significant deviation compared to the RWA solutions [black lines]. In this regime of ultrastrong coupling, the ground state and the polaritons [indicated by arrows in Fig.~\ref{fig:spec}(a)] are significantly dressed by higher excitations, as will be corroborated in Sec.~\ref{sec:Ncav_GS}. We note that $\tilde{g}_\text{th}^\text{US}$ is insensitive to the amount of disorder, while $\tilde{g}_\text{th}^\text{pol}$ depends on the disorder strength. Moreover, as shown below, the polariton formation is subject to severe finite size effects in the presence of disorder. Thus, $\tilde{g}_\text{th}^\text{pol}$ can be larger than $\tilde{g}_\text{th}^\text{US}$ when disorder is extremely large. 

At still stronger couplings of $\tilde{g} \gtrsim \omega_\text{c}/2$,  states with different excitation numbers $M$ show level crossings. There are no avoided crossings between parity sectors $M(\mathrm{mod} 2)$, because these represent irreducible representations of the parity symmetry group. In particular the ground state and the dressed lower polariton (DLP) become nearly degenerate, which, even at $L=6$, indicates that parity symmetry is broken, and the system enters the superradiant phase. Indeed, the mean field approximation yields the critical point at exactly $\tilde{g}=\tilde{g}_\text{c}^\text{MF,SR}=\omega_\text{c}/2$, as we demonstrate in Appendix~\ref{app:MF}. The true transition should naturally lie at a higher value of $\tilde{g}$, since mean field theory tends to overestimate the extent of the broken symmetry (large $g$-)phase. For the remainder of the paper, we shall not focus on the superradiant transition. On the one hand, it is beyond the coupling strengths achieved in organic materials. On the other hand, an all-to-all dipole coupling term becomes relevant in this regime, and may suppress the transition~\cite{debernardisCavityQuantumElectrodynamics2018}.

The main effect of doping on the spectrum is to increase the number of states. The ground state as well as polaritons [indicated by arrows in Fig.~\ref{fig:spec}(a)] are replaced by $L$ nearly degenerate states forming bands [see Fig.~\ref{fig:spec}(b)]. All of these bands appear to have a $g$-independent bandwidth, which we confirm below in Sec.~\ref{sec:LDOS_el}. This last observation suggests that conduction electrons are unaffected by the light-matter coupling. Indeed, this is plausible, because doubly occupied and completely empty lattice sites do not couple to the cavity.

Finally, the disorder-averaged spectra [Fig.~\ref{fig:spec}(c) and (d)] fulfill all the same properties as described above. The difference compared to the homogeneous cases is that now the bandwidth $W$ is not determined by the hopping $J_2$ but by the width of distributions of $\delta J_2$ and $\delta \epsilon_2$. As we show below, the disorder changes the properties of the states between the UP and LP such that all of them appear in the one-photon excitation spectrum. By contrast, in the homogeneous case only a subset of these states couple to the cavity.

\begin{figure}
	\centering
	\includegraphics[width=\columnwidth]{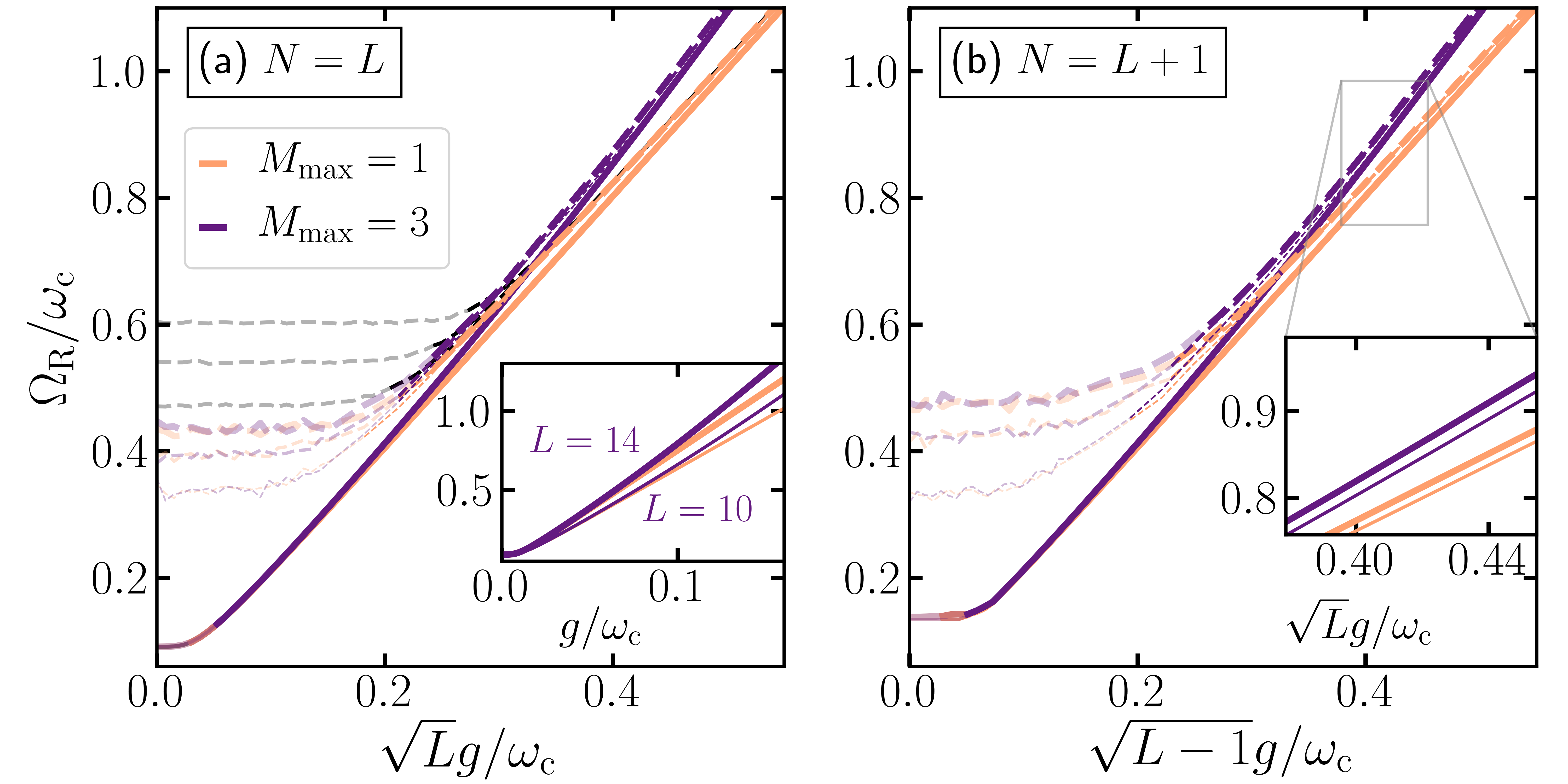}
	\caption{Vacuum Rabi splitting $\Omega_\text{R}$ as a function of the rescaled light-matter coupling in an undoped system (a) and with a single conduction electron (b). A comparison is made between the homogeneous (full lines) and disorder-averaged (dashed lines) case, between the RWA (orange) and corrections involving $M=3$ excitations (purple), and between $L=6,10,14$ (increasing line width). For the undoped case with the RWA, we additionally show results up to $L=80$ (black).}
	\label{fig:Rabi}
\end{figure}

\begin{figure}
	\centering
	\includegraphics[width=\columnwidth]{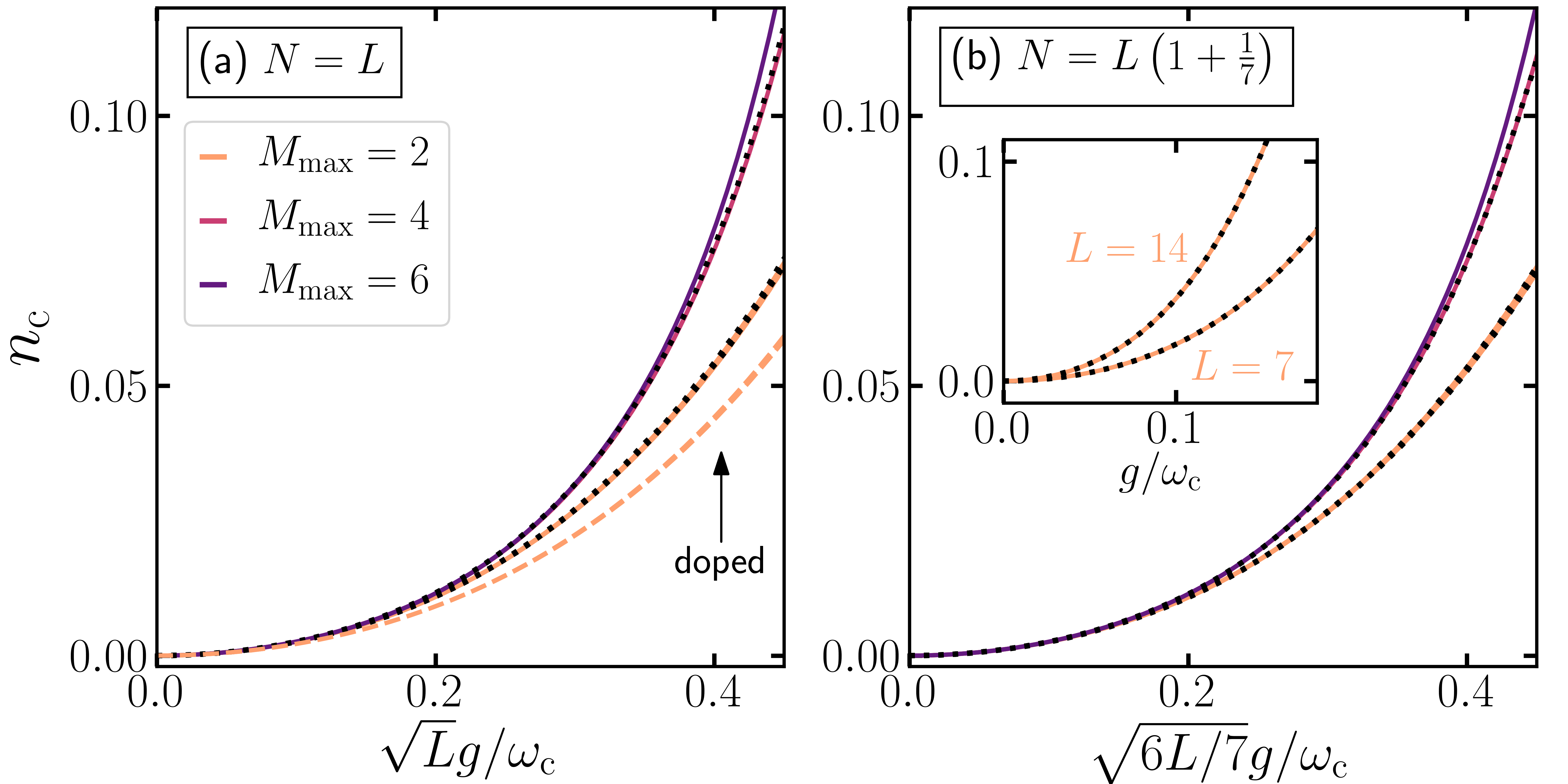}
	\caption{Cavity occupation number in the ground state for (a) the undoped and (b) the doped system with doping density $n_\text{d}=1/7$. One of the doped cases is also shown in (a) for comparison (dashed line). Corrections up to $M=6$ (purple) are taken into account and compared to $M=4,2$ (red, orange). The system size is $L=7$ (all cases) and $L=14$ (only for $M_\text{max}=2$). Black dotted lines correspond to the disordered cases. The inset of (b) shows the case $M_\text{max}=2$ against the microscopic coupling strength $g/\omega_\text{c}$.}
	\label{fig:ncav}
\end{figure}

\subsection{Vacuum Rabi splitting}\label{sec:Rabi}

In order to assess more thoroughly the scaling of the spectrum with $g$ and $L$, the vacuum Rabi splitting $\Omega_\text{R}$, i.e., the difference in energy of the upper and lower polaritons is depicted in Fig~\ref{fig:Rabi}. We compute the lowest and highest energy of the ($M=1$)-sector using the Lanczos method. 

At weak coupling, whether doped or not, there is always a region where the Rabi splitting is apparently constant (blurred lines in Fig.~\ref{fig:Rabi}). This regime is determined by the furthest outliers of the energies of the fermionic part of the model. Here, the Rabi splitting cannot be properly defined, because there is no clear separation of the polaritons from the other energy levels. We note that, in the disordered case, the expectation value of the difference between these outliers increases with the system size [see especially Fig.~\ref{fig:Rabi}(a)]. As we will see in Sec.~\ref{sec:LDOS_exc}, this leads to severe finite size effects regarding the formation of idealized exciton-polaritons as given in Eq.~\eqref{eq:polariton}. In real materials, $L$ does not correspond to the sample size, but to the coherence length of exciton-polaritons. Nevertheless, our results show that a large static disorder can be a hindrance to exciton-polariton formation.

On the other hand, for sufficiently strong coupling, we see the disordered and homogeneous cases collapse onto the same trajectories. Corrections by higher excitation numbers become relevant around $g\sqrt{L(1-n_\text{d})}\gtrsim \tilde{g}_\text{th}^\text{US} \approx 0.2\omega_\text{c}$, as expected, but they do not affect the results described here.

A comparison of different system sizes confirms our scaling assumption. For $L=10$ (thin) and $L=14$ (thick), the lines are on top of each other, when plotted against the rescaled coupling $\tilde{g}=g\sqrt{L}$ for the undoped case in Fig.~\ref{fig:Rabi}~(a), and $\tilde{g}=g\sqrt{L-1}$ for the case with a single conduction electron in Fig.~\ref{fig:Rabi}~(b). The insets show that this is not the case with other rescaling factors.

\subsection{Cavity population of the dressed vacuum}\label{sec:Ncav_GS}

Here, we quantify the dressing of eigenstates by the counter-rotating terms in the ultrastrong coupling regime. We show the cavity occupation number $n_\text{c}=\braket{0|a^\dagger a|0}$ in Fig.~\ref{fig:ncav}, where $\ket{0}$ depends on $g$. To check convergence, $M=2,4,6$ are compared. Indeed, up to $\tilde{g}\sim 0.3\omega_\text{c}$ it is sufficient, to take into account up to $M_\text{max}=2$ excitations, whereas $M_\text{max}=4$ is sufficient up to $\tilde{g}\sim 0.4\omega_\text{c}$. The former corresponds to first-order perturbation theory in $\tilde{g}$ for the ground state, and therefore to a correction $\sim\tilde{g}^2$ of the photon number. 

Disorder has no influence the photon content of the ground state (black dotted lines in Fig.~\ref{fig:ncav}). Indeed, the excitation gap from the noninteracting vacuum $\ket{\text{FS}}\otimes \ket{0}$ to the $(M=2)$-sector is approximately $2\omega_\text{c}$, i.e., far away from resonance. Hence, the fluctuations of the molecular energy levels have only negligible influence. Meanwhile, resonant effects, such as the formation of exciton-polaritons that occur within the $(M=1)$ sector, are very sensitive to these fluctuations. 

We further find that the photon number scales with the combination $\tilde{g}=g\sqrt{L(1-n_\text{d})}$, as well, since the undoped case in Fig.~\ref{fig:ncav}(a) behaves exactly as the doped case with $n_\text{d}=1/7$ in Fig.~\ref{fig:ncav}(b). This means that the doping reduces the effective coupling strength [see dashed line in panel (a)].

\section{Local densities of state}\label{sec:LDOS}

So far, we have considered only the energy eigenvalues and the ground state photon number. However, the contributions of various physical excitations to these eigenstates are nontrivial. Besides cavity photons and conduction electrons and holes, there are excitations like the bright mode created by the operator in Eq.~\eqref{eq:bright_mode} and other electron-hole (e-h) excitations. In order to quantify these contributions, one can define various local densities of states (LDOS). 

The general form of the LDOS derives from an autocorrelation function in frequency space of a local operator $A$~\cite{bruusManybodyQuantumTheory2004}, 
\begin{equation}
	\begin{split}
		G_A(\omega+i\gamma) &= -i \int_{0}^{\infty} \braket{ A^\dagger(t) A} e^{i(\omega+i\gamma)t} \text{d}t \\
		&= -i \int_{0}^{\infty} \braket{ e^{iHt} A^\dagger e^{-iHt} A} e^{i(\omega+i\gamma)t} \text{d}t,
	\end{split}
\end{equation}
where we have added a finite broadening $\gamma$ for convergence. While the expectation value can be taken with respect to any state, presently the ground state is a good choice, i.e., $\braket{\ldots}=\braket{0|\ldots|0}$, such that
\begin{equation}\label{eq:G_A}
	G_A(\omega+i\gamma) = \braket{0| A^\dagger \frac{1}{ \omega + i\gamma - H + E_0 } A |0},
\end{equation}
where $E_0=\braket{0|H|0}$ is the energy of the ground state. The LDOS is defined as the imaginary part thereof, along with a normalization factor,
\begin{equation}\label{eq:rho_A}
	\rho_A(\omega+i\gamma) = -\frac{1}{\pi} \Im G_A(\omega+i\gamma).
\end{equation}
In order to compute the LDOS, we employ the Lanczos algorithm for the ground state and a continued fraction expansion for finding the projection of the state $(\omega + i\gamma - H + E_0 )^{-1} A \ket{0}$ onto $A\ket{0}$~\cite{gaglianoDynamicalPropertiesQuantum1987,gruningImplementationTestingLanczosbased2011}.

In this section, we focus on the half-filling case, $N=L$. The ground state is the dressed vacuum [see Fig.~\ref{fig:ncav}] of the interacting system of fermions and cavity. Starting from this ground state, we consider the LDOS of the cavity,
\begin{equation}\label{eq:rho_a}
	\rho^\text{cav} = \rho_{a^\dagger} 
\end{equation}
of electrons and holes at site $j$,
\begin{equation}\label{eq:rho_el}
	\begin{split}
		\rho_{j}^\text{el} &= \rho_{c_{2,j}^\dagger} \\ 
		\rho_{j}^\text{h} &= \rho_{c_{1,j}} 
	\end{split}
\end{equation}
and for e-h excitations,
\begin{equation}\label{eq:rho_ex}
	\begin{split}
		\rho_{j,r}^\text{eh} 
		= \rho_{ c_{2,j+r}^\dagger c_{1,j} }. 
	\end{split}
\end{equation}

There are two coordinates for an e-h excitations, which we cast as the center of mass (COM) $j$ and the relative coordinate $r$ of the electron-hole pair. Due to the periodic boundary condition, it is permissible to choose the COM as the position of the hole, which simplifies notation. Further insight can be gained by considering a coarse-grained LDOS for each of these coordinates by tracing out the other one,
\begin{equation}\label{eq:rho_ex-com-rel}
	\begin{split}
		\rho_{j}^\text{COM} &= \sum_r \rho_{j,r}^\text{eh}, \\
		\rho_{r}^\text{rel} &= \sum_j \rho_{j,r}^\text{eh}.
	\end{split}
\end{equation}

Finally, we can define the total density of states (DOS) for electrons, holes and e-h excitations, by summing over all coordinates, i.e.,
\begin{equation}
	\begin{split}\label{eq:DOS_ex-com-rel}
		\text{DOS}^\text{el/h} = \sum_j \rho_{j}^\text{el/h} \\
		\text{DOS}^\text{eh} = \sum_{j,r} \rho_{j,r}^\text{eh}.
	\end{split}
\end{equation}
We note that these are still more sparse than the total density of states of the Hamiltonian, $D(\omega)=\sum_n \delta(E_n-\omega)$ with spectrum $\{E_n\}$, because some states may have little or no weight with respect to the excitations in Eq.~\eqref{eq:DOS_ex-com-rel}.

\subsection{Photonic spectral function}\label{sec:LDOS_cav}

\begin{figure}[t!]
	\centering
	\includegraphics[width=\columnwidth]{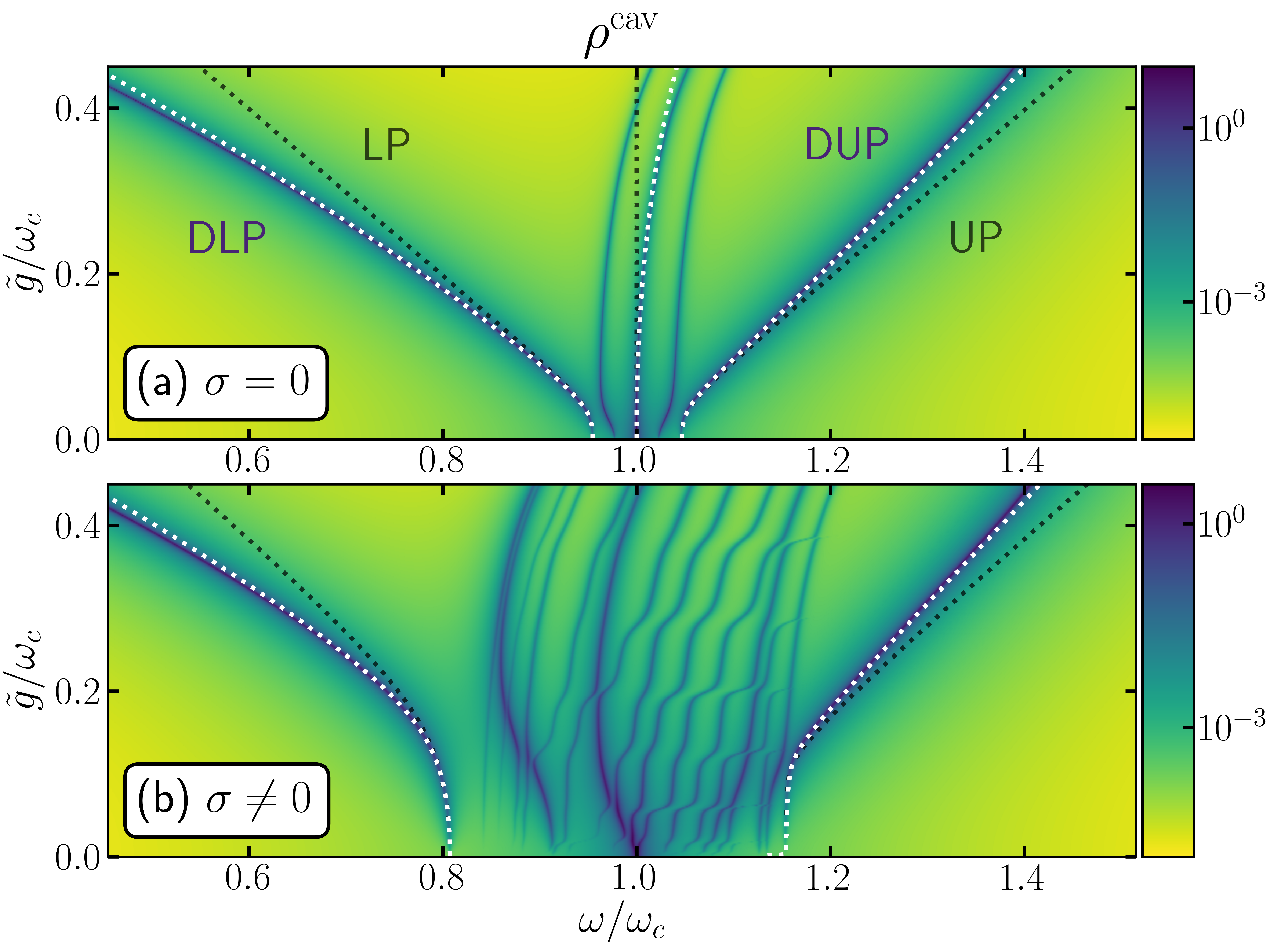}
	\caption{Log-scale Cavity LDOS $\rho^\text{cav}$ for (a) the homogeneous case and (b) a single disorder realization. The system size is $L=6$, and up to $M=5$ excitations are taken into account. White (gray) dotted lines indicate the energy of the dressed UP and LP for $M=3$ ($M=1$), as well as the center of the region of in-gap states in panel (a).}
	\label{fig:LDOS_cav_2d}
\end{figure}

The LDOS of the cavity, $\rho^\text{cav}$, is depicted in Fig.~\ref{fig:LDOS_cav_2d} as a function of the coupling $\tilde{g}=g\sqrt{L}$ and frequency $\omega$ for (a) the homogeneous case and (b) for the disordered case with up to $M=5$ excitations. Here, as in the rest of this section, the broadening $\gamma$ is chosen such that individual states are resolved. Even for relatively large couplings, the DLP/DUP for $M=5$ and $M=3$ (white dotted lines) are very close, indicating the convergence of or calculations in terms of $M$.

\begin{figure*}[t!]
	\centering
	\includegraphics[width=\textwidth]{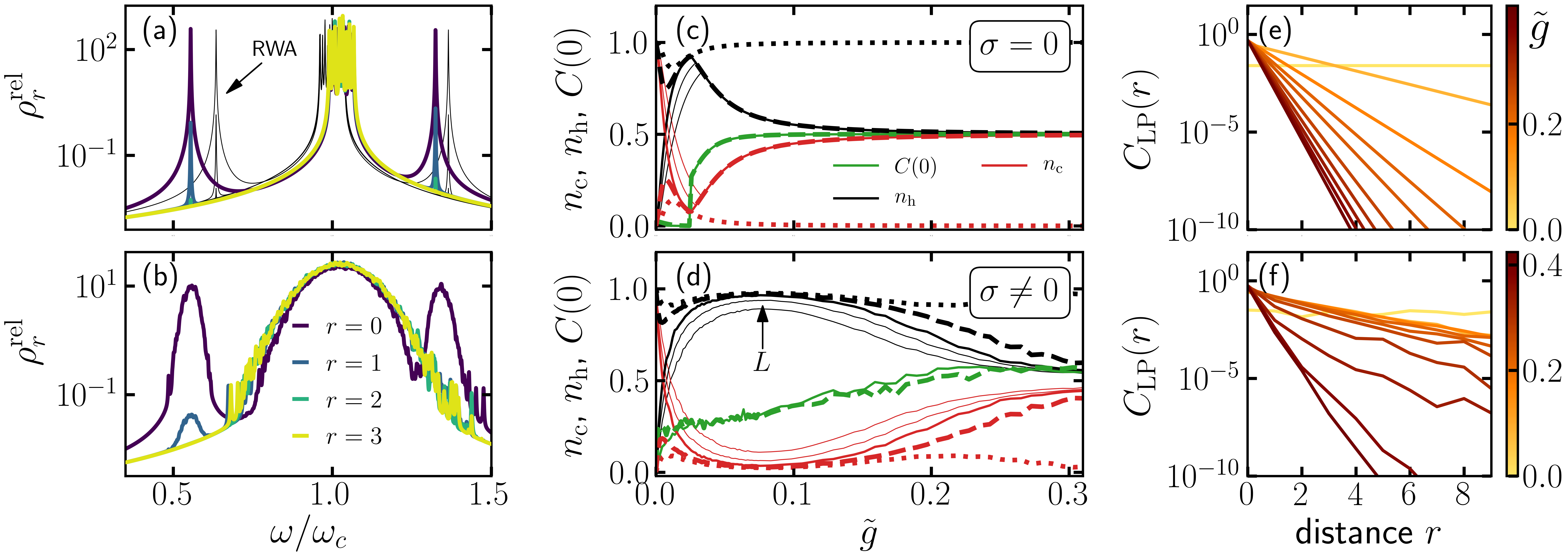}
	\caption{Upper panels (a),~(c)~and~(e) show data for the homogeneous case, lower panels (b),~(d)~and~(f) show data for the disordered case. The disorder averages in the bottom row are obtained from $2000$ realizations. 
	Left panels [(a) and (b)]: Exciton LDOS $\rho_r^\text{rel}$ with relative coordinate $r=0,1,2,3$ at $\tilde{g}/\omega_\text{c}=0.3$ and $N=L=6$. Up to $M=5$ excitations are taken into account. Thin black lines represent the RWA with $M=1$.
	Middle panels [(c) and (d)]: electron-hole correlation function $C(0)$, number of holes $n_\mathrm{h}$, and number of photons $n_\mathrm{c}$ are plotted for the states of the largest (solid lines), second largest (dashed lines) and third largest (dotted lines) number of photons. The system size is $L=79$, while for the first state $L=19,39$ are shown in addition (thin lines). 
	On the right [(e) and (f)], the electron-hole correlation functions $C(r)=C_\text{LP}(r)$ of the second state are shown as a function of the relative coordinate $r$ for $L=39$ for various values of $\tilde{g}$ indicated by the color bar.}
	\label{fig:eh_corr_LDOS}
\end{figure*}

Interestingly, the cavity LDOS in Fig.~\ref{fig:LDOS_cav_2d}(a) indicates that, even at vanishing disorder, there exist in-gap states with significant photon contents up to $\tilde{g}\sim 0.1\omega_\text{c}$, which is not clear from the mere spectrum of Fig.~\ref{fig:spec}. This is in contrast to pure exciton or spin-half models, like the Dicke model, where these in-gap states are completely dark for vanishing disorder. The reason for these bright in-gap states is that the dispersion relations of the conduction and valence bands are not the same, i.e., $J_1\neq J_2$. As a consequence, the bright mode operator of Eqs.~\eqref{eq:bright_mode} creates a superposition of nondegenerate eigenstates $b_{2,k}^\dagger b_{1,k}\ket{FS}$ of $H_\text{el}$ with energies $2(J_2-J_1)\cos(k) + \epsilon_2-\epsilon_1$. Only in the limit where the coupling is much larger than the spread of these energies, i.e., $\tilde{g} \gg |J_2-J_1|$, can we approximately describe the polaritons in terms of a single matter excitation, as in Eq.~\eqref{eq:polariton}. Conversely, the in-gap states have a significant overlap with the bright mode at intermediate coupling strengths.

The distinctive feature of $\rho^\text{cav}$ in the disordered system, shown in Fig.~\ref{fig:ncav}(b) for a single disorder realization, is a larger number of peaks. Indeed, if the bright mode operator is expressed in the electronic eigenstates of the disordered Hamiltonian, one obtains a superposition of $L^2$ instead of $L$ states, i.e.,
\begin{equation}
	B^\dagger = \sum_{n,m} B_{nm} d_{2,n}^\dagger d_{1,m}.
\end{equation}
Here $d_{\nu,m} = \sum_j U_{mr}^{(\nu)} c_{\nu,r}$ is an eigenstate and the coefficient for the coupling between states in the lower and upper band is given by $B_{nm}=\sum_{r,s} \left(U_{mr}^{(2)}\right)^\dagger U_{ns}^{(1)}$. 

In principle, each of these excitations contributes to $\rho^\text{cav}$. However, only $\sim L$ of the peaks in Fig.~\ref{fig:LDOS_cav_2d}(b) are pronounced. We can understand this as a consequence of Anderson localization: there is an approximate correspondence between the tight-binding orbitals created by $c_{\nu,r}^\dagger$ and the present eigenstates. With appropriately ordered indices, the elements of the basis transformation matrix are exponentially bounded, i.e., $|U_{mr}|\sim \exp{\left(-\xi|m-r|\right)}$ with localization length $\xi$. Since the bright mode contains only e-h excitations at relative distance $r=0$, it overlaps mainly with such pairs of electron and hole eigenstates. The results in Fig.~\ref{fig:LDOS_cav_2d}(b) suggest that the localization length $\xi$ is on the order of the lattice spacing. Finally, we remark that the bright in-gap states may be related to the semilocalized dark states discussed in Ref.~\cite{botzungDarkStateSemilocalization2020}.

\subsection{Exciton density of states and e-h correlations}\label{sec:LDOS_exc}

To expand on the previous analysis, we show the LDOS for e-h excitations $\rho_r^\text{rel}$ for different relative distances $r$ at $\tilde{g}=0.3\omega_\text{c}$ in Fig.~\ref{fig:eh_corr_LDOS}(a)~and~(b). At the polariton peaks, there are clearly contributions from nonzero distances $r\neq0$. However, they fall off exponentially as $\rho_r^\text{rel}\sim e^{-r/\ell}$, which is especially striking in the polariton peaks corresponding to the homogeneous case in Fig.~\ref{fig:eh_corr_LDOS}(a). This suggests that the polariton-formation is accompanied by the formation of excitons characterized by a Bohr radius $\ell$. Furthermore, the RWA [black lines in Fig.~\ref{fig:eh_corr_LDOS}(a)] merely shifts the peak position, whereas it does not affect the exponential relation. Thus, exciton-polariton formation is a resonant phenomenon.

We make the notion of exciton formation more precise by explicitly considering the electron-hole correlation function
\begin{equation}
	C(r) = \sum_j \braket{(1-n_{1,j})n_{2,j+r}},
\end{equation}
see Fig.~\ref{fig:eh_corr_LDOS}(c)-(e). Here, we restrict ourselves to the single-excitation sector ($M = 1$) by using the RWA, as is justified by Fig.~\ref{fig:eh_corr_LDOS}(a). This allows the full diagonalization even at large system sizes $L\sim 100$. Moreover, we compute the photon numbers $n_\text{c}$ for each of the obtained eigenstates. By choosing the states with the three largest values of $n_\text{c}$, we pick up the two polaritons and one additional state, even when they are still hidden in terms of the energy spectrum. Indeed, in the disordered case, there is always a chance to find an outlier of energy $E<E_\text{LP}$ or $E>E_\text{UP}$.

In Fig.~\ref{fig:eh_corr_LDOS}(c)~and~(d), we show the e-h correlation function $C(0)$ at relative distance $r=0$ (green lines) with the photon number $n_\mathrm{c}$ (red lines) and the number of holes $n_\mathrm{h}=\sum_j (1-n_{1,j})$ (black lines), where $n_{\nu,j}=\braket{ c_{\nu,j}^\dagger c_{\nu,j}}$. By choice of the undoped subspace with $N=L$, the number of conduction electrons is equal to the number of holes, $n_\text{e}=n_\text{h}$. The restriction to $M=1$ implies $0\leq n_\text{e/h}\leq 1$.

In all the cases the state of the highest photon number (solid lines) starts from $n_\text{c}=1$ and $n_\text{h} = 0$ at $\tilde{g}=0$, where it is given by $\ket{\text{FS}}\otimes\ket{1}$. Meanwhile, the second state (dashed lines) and third state (dotted lines) start from $n_\text{c}=0$ and $n_\text{h} = 1$. As shown in Fig.~\ref{fig:eh_corr_LDOS}(e)~and~(f) for the second state, the hole is uncorrelated with the conduction electron [$C(0) \approx 1/L$] at $\tilde{g}=0$. For the homogeneous case, this is a property of the two-body Bloch states $b_{2,k}^\dagger b_{1,k'}\ket{\text{FS}}$. In the disordered system, on the other hand, it is a consequence of the averaging over disorder realizations. Concerning the opposite limit of large coupling, we find that the first and second state can be clearly identified as the UP and LP with nearly equal hole and photon numbers ($n_\text{h}\approx n_\text{c} \approx 1/2$). Here, electrons and holes are almost perfectly correlated $C(0)/\sqrt{n_\text{h}n_\text{e}} \approx 1$. In this strong-coupling regime, the correlation function $C(r)$ indeed shows an exponential decay [see Fig.~\ref{fig:eh_corr_LDOS}(e)~and~(f)], as expected from the behavior of the LDOS $\rho_r^\text{rel}$. These spatially correlated electrons and holes can be regarded as excitons.

Between the limits, there is a transition region with nonanalyticities of the observables as a clear signature of level crossings in the homogeneous case in Fig.~\ref{fig:eh_corr_LDOS}(c). Such features are washed out by the disorder averaging [see Fig.~\ref{fig:eh_corr_LDOS}(d)]. In both cases, however, there is a certain coupling strength $\tilde{g} = \tilde{g}^\text{macro} < g_\text{th}^\text{pol}$ where the cavity excitation is distributed almost equally among a macroscopic number of states; note the increasing system sizes shown for the state of the largest photon number (thin solid lines), which indicate $n_\text{c}\to 0,\,n_\text{h}\to 1$ for $L\to\infty$. Whether this point marks an actual phase transition is an open question. 

In the subsequent region of polariton formation, $\tilde{g}^\text{macro} < \tilde{g} \lesssim g_\text{th}^\text{pol}$, where the photon and hole numbers are still significantly below the value of $1/2$, the e-h correlation function is already close to $1/2$. Thus, the exciton formation precedes the exciton-polariton formation. Moreover, when disorder is present, the polariton formation seems to shift to larger couplings as $L$ is increased, $\tilde{g}^\text{pol}=\tilde{g}^\text{pol}(L)$. This is a consequence of outliers in the electronic spectrum as found in Sec.~\ref{sec:Rabi}. Both of these effects induce features in the inverse participation ratios presented in Sec.~\ref{sec:GIPR}.

\begin{figure}
	\centering
	\includegraphics[width=\columnwidth]{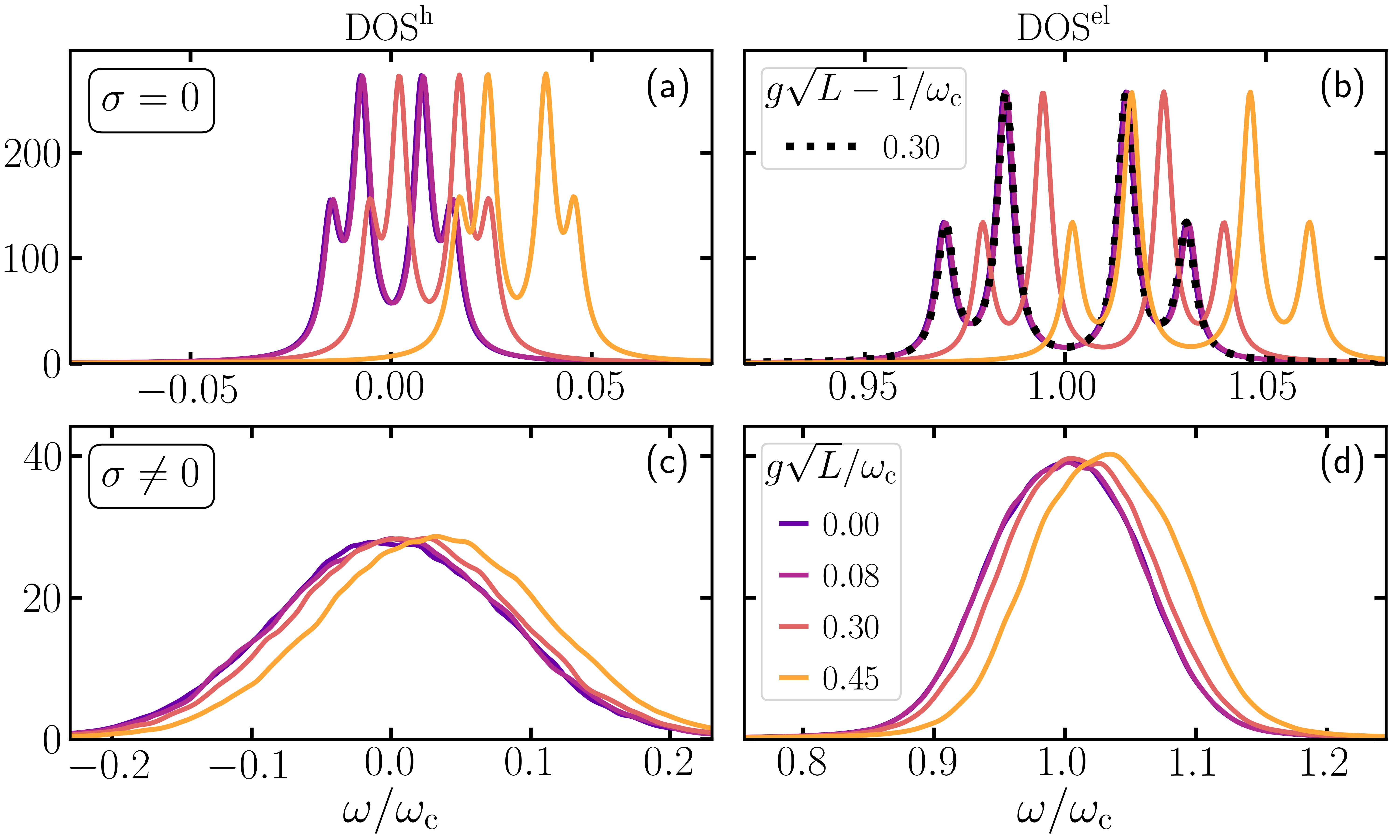}
	\caption{DOS for holes (left) and electrons (right) for various values of the coupling $\sqrt{L}g$ at $L=6$ with up to $M=6$ excitations. Panels (a) and (b) show the homogeneous case, panels (c) and (d) show the disordered case. The black dotted line represent the DOS at a large coupling strength where the ground state $\ket{0}$ is computed for a certain effective value $\tilde{g}=\sqrt{L}g$, but a rescaled effective value $\tilde{g}^\prime=\sqrt{L-1}g$ is used for the computation of the DOS.}
	\label{fig:DOS_el_h}
\end{figure}

\subsection{Electron and hole density of states}\label{sec:LDOS_el}

Besides excitons and cavity photons, which are coupled by $H_\text{el-cav}$, one may wonder how individual electrons and holes are affected by the cavity. In the original experimental work on cavity-enhanced charge transport~\cite{orgiuConductivityOrganicSemiconductors2015}, it was suggested that electronic states delocalize under strong light-matter coupling. Another work~\cite{hagenmullerCavityEnhancedTransportCharge2017}, proposed a mechanism where the mobility of holes increases due to an additional hopping channel through the upper band. While we address the former suggestion in the following section, we now discuss the latter in the framework of the electron- and hole-DOS.

The DOS of the charged degrees of freedom in homogeneous and disordered systems are shown in Fig.~\ref{fig:DOS_el_h}. Differences in electronic and hole bandwidths are, respectively, due to different hopping parameters $J_\nu$ and standard deviations of the random variables $\delta J_\nu$ and $\delta \epsilon_\nu$. The main effect of the light-matter coupling is a shift in energy relative to the ground state. However, this is solely due to the fact the microscopic coupling strength $g$ is the same for the computation of the ground states, where $\tilde{g}=g\sqrt{L}$, and for the LDOS $\rho^{\text{el/h}}$, where $\tilde{g}=g\sqrt{L-1}$ because of the charge. If $g$ is rescaled appropriately, the energy shift vanishes [see black dotted line in Fig.~\ref{fig:DOS_el_h}(b)].

More importantly, the bandwidths are not affected by the light-matter coupling. This indicates that, within the present model of a closed system, there is no increase in charge mobilities. We come back to this in Sec.~\ref{sec:GIPR}. Furthermore, bandwidths of electrons and holes are affected in the same manner, which implies that an extra hopping channel for holes cannot be at play here, in contrast to the open system studied in Ref.~\cite{hagenmullerCavityEnhancedTransportCharge2017}.

\section{Localization}\label{sec:GIPR}
Having characterized the eigenstates in terms of excitations of various degrees of freedom, we now discuss the conduction properties in terms of the localization of these excitations. As a measure of localization, we employ the generalized inverse participation ratio (GIPR)~\cite{murphyGeneralizedInverseParticipation2011} based on the (possibly coarse-grained) LDOS introduced above,
\begin{equation}
	G(\omega+i\gamma) = \frac{\sum_j \rho_j(\omega+i\gamma)^2}{\left(\sum_j \rho_j (\omega+i\gamma)\right)^2}. 
\end{equation}
The denominator ensures normalization, such that cases with different total DOS can be compared, whereas the numerator measures how states close to an energy $\omega$ are distributed over a set of local excitations. 

The footprint of localization is that the GIPR is constant under finite size scaling. Indeed, in the totally localized limit, we have $\rho_j \sim \delta_{j,j_0}$, whereas in the delocalized limit we have $\rho_j \sim 1/L$ for all $j$, supposing the summation includes $L$ terms. For the GIPR, it follows that
\begin{equation}
	G(L) \sim 
	\begin{cases}
		\text{const.} & \text{(localized)}\\
		1/L & \text{(delocalized/extended)}
	\end{cases}.
\end{equation}

\begin{figure}
	\centering
	\includegraphics[width=\columnwidth]{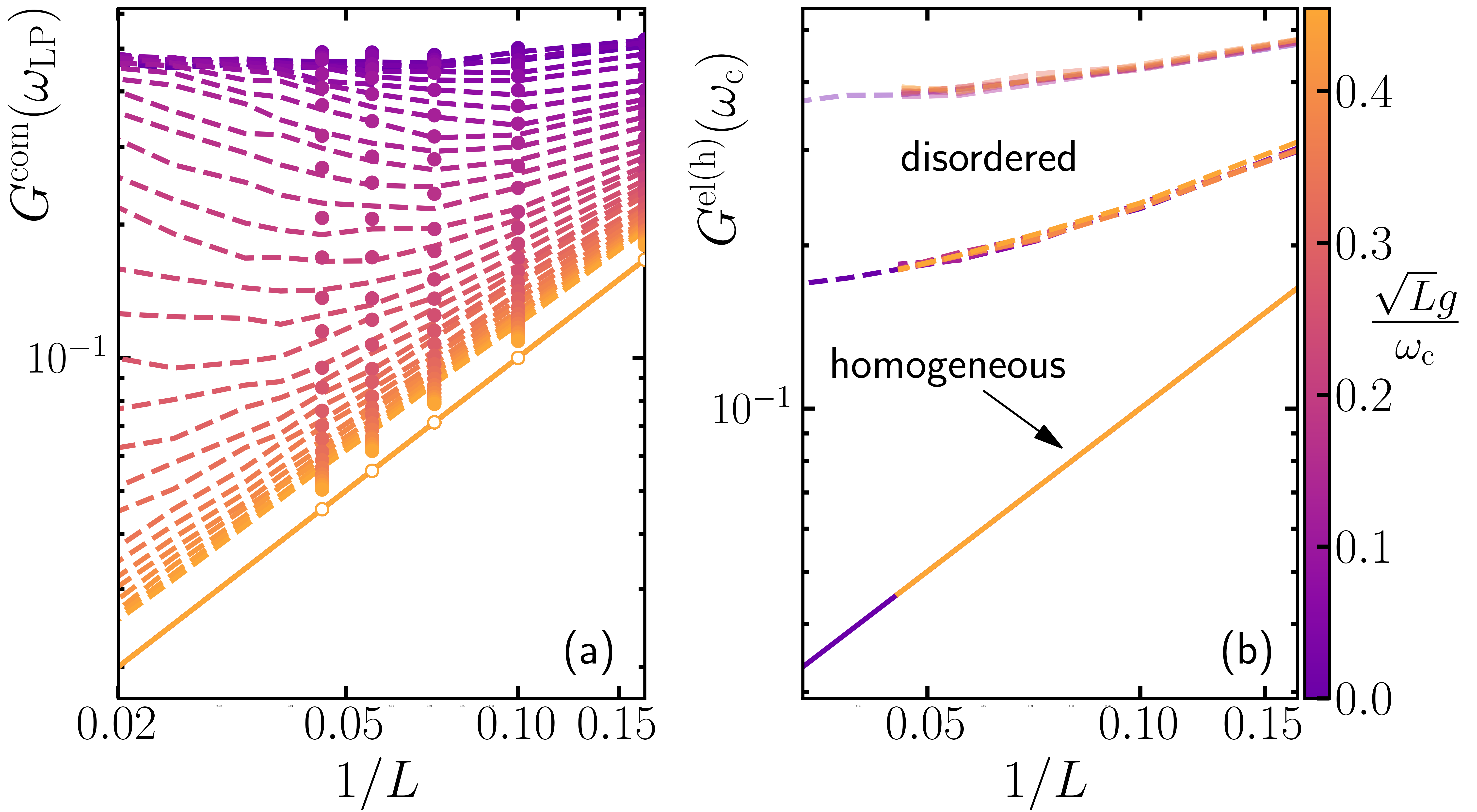}
	\caption{GIPR for (a) the COM of e-h excitations at the LP energy $\omega_\text{LP}$ as extracted from Fig.~\ref{fig:LDOS_cav_2d}, and (b) for electrons (nontransparent lines) and holes (transparent lines). Homogeneous (solid lines) and disordered (dashed lines) cases are compared. In (a) we include up to $M=3$ (dots) or $M=1$ (lines) excitations, and in (b) up to $M=2$ excitations.}
	\label{fig:GIPR}
\end{figure}

Here, as shown in Fig.~\ref{fig:GIPR}, we consider electrons, holes, and e-h excitations with COM $j$,
\begin{equation}
	\begin{split}
		G^\text{el/h} &= \frac{\sum_j \left(\rho_j^\text{el/h}\right)^2}{\left(\text{DOS}^\text{el/h}\right)^2} \\
		G^\text{COM} &= \frac{\sum_j \left(\rho_j^\text{COM}\right)^2}{\left(\text{DOS}^\text{COM}\right)^2}.
	\end{split}
\end{equation}
More specifically, for electrons, we chose the frequency as the cavity frequency $\omega=\omega_\text{c}$, which corresponds to states in the upper band, as shown by the electronic DOS in Fig.~\ref{fig:DOS_el_h}(b)~and~(d). Similarly, for holes, the frequency is $\omega=0$. Finally, for e-h excitations, we are especially interested in the lower polariton. Therefore, the choice of a frequency depends on the coupling strength, i.e., $\omega_\text{LP}(\tilde{g})$, which we extract from Fig.~\ref{fig:LDOS_cav_2d}. 

Furthermore, the broadening $\gamma$, needs to be adjusted depending on the system size. The subtle details of this issue have been elaborated in Ref.~\cite{murphyGeneralizedInverseParticipation2011} for a simple model system. Following these results, we chose $\gamma\sim 1/L$ for electrons and holes, and $\gamma\sim1/L^2$ for e-h excitations.

First, we discuss the case of e-h excitations, which is depicted in Fig.~\ref{fig:GIPR}(a). At vanishing coupling (dark purple), disorder leads to localization while excitations in the homogeneous system are delocalized. As the coupling increases, the effect of disorder is lifted (yellow), and the states become delocalized. This observation is in line with previous studies that found enhanced exciton mobilities under strong light-matter coupling~\cite{schachenmayerCavityEnhancedTransportExcitons2015,feistExtraordinaryExcitonConductance2015}. Here, we further find that effects beyond the RWA play no essential role [see dots in Fig.~\ref{fig:GIPR}(a)].

In addition, there is an intermediate regime of coupling strengths, where we see $G^\mathrm{COM}(\omega_\mathrm{LP})\sim 1/L$ for small $L$, while the GIPR increases again for larger $L$. This fact is explained in terms of Fig.~\ref{fig:eh_corr_LDOS}(d), where we found a system-size-dependent threshold value $\tilde{g}_\text{th}^\text{pol}(L)$ for exciton and exciton-polariton formation, i.e., for the strong coupling regime; The GIPR falls off, while $\tilde{g} > \tilde{g}_\text{th}^\text{pol}(L)$. However, this trend reverses when $\tilde{g} < \tilde{g}_\text{th}^\text{pol}(L)$, where the $L$-dependent disorder is so large that the system is no longer in the regime of strong coupling. 

Second, for individual charges [see Fig.~\ref{fig:GIPR}(b)], we find that strong coupling has no effect on the localization. Electrons (nontransparent lines) as well as holes (transparent lines) remain localized in the presence of disorder. The merely quantitative difference between the GIPR of electrons and holes is due to our choice of larger disorder fields in the lower band compared to the upper band. This observation is in agreement with the fact that bandwidths of electrons and holes are unaffected by strong coupling [see Fig.~\ref{fig:DOS_el_h}]. Thus, while the states can be significantly dressed by photons and excitons at large coupling, the (de-)localized charges are essentially independent of such excitations. The only possible effect of exciton-dressing is a decreasing charge mobility due to the Pauli principle. However, these interactions become irrelevant in our model at large $L$, because the density of excitons decreases with $1/L$.

\section{Conclusion}\label{sec:conclusion}
In this paper, we have investigated the relationship between the excitation spectrum and localization of disordered two-band electrons coupled to a cavity mode. We have shown that the model is a close analog of the Dicke model where electron and hole doping modify the scaling form of the energy spectrum. The transition towards strong coupling features a point where the cavity excitation is shared by a macroscopic number of eigenstates. In contrast to previous studies indicating the delocalization of electronic wavefunctions, the inverse participation ratios show that electrons and holes are still localized. However, we also find that excitons localized by disorder can be delocalized due to the light-matter coupling, which may enhance the electronic conductivity indirectly in combination with electrodes. Our results provide insight into the transport behavior of strongly coupled systems and suggest a potential route towards designing efficient solid-state devices.

\hspace{10pt}

\begin{acknowledgments}
	The authors would like to thank Jianshu Cao and Michael Thoss for fruitful discussions. We are grateful for the support by the state of Baden-W\"urttemberg through bwHPC and the German Research Foundation (DFG) through grant no INST 40/575-1 FUGG (JUSTUS 2 cluster). JO also acknowledges support from the Georg H. Endress Foundation. 
\end{acknowledgments}

\appendix
\onecolumngrid
\section{Mean-field theory and superradiant transition}\label{app:MF}
A mean-field theory of our model can be developed in analogy to the Dicke model. Assuming that the cavity is in a coherent state $\ket{\alpha}$, defined by $a\ket{\alpha}=\alpha\ket{\alpha}$, a parametrized family of Hamiltonians for the electrons is obtained,
\begin{equation} \label{app:eq:H_MF}
	H(\alpha) = H_\text{el} + \omega_\text{c} |\alpha|^2 + g (\alpha + \alpha^\ast) \sum_r \left[ c_{2,r}^\dagger c_{1,r} + c_{1,r}^\dagger c_{2,r} \right],
\end{equation}
which leads to the respective partition functions $Z(\alpha)$ and free energies $F(\alpha)$. In the following, $\alpha$ is assumed to be real.

In the Dicke model, a superradiant phase is defined such that the photon number increases faster than the square root of the system size~\cite{heppSuperradiantPhaseTransition1973a,wangPhaseTransitionDicke1973a}. Thus, one introduces a rescaled order parameter, as well as a rescaled coupling strength,
\begin{equation}\label{app:eq:rescaling}
	\tilde{\alpha} = \frac{\alpha}{\sqrt{L}}, \qquad \tilde{g} = \sqrt{L}g.
\end{equation}
The critical point $\tilde{g}_\text{c}$ is defined by
\begin{equation}
		\lim_{L\to\infty} \tilde{\alpha}
		\begin{cases}
			= 0  \qquad \text{for } \tilde{g} \leq \tilde{g}_\text{c} \\
			> 0  \qquad \text{for } \tilde{g} > \tilde{g}_\text{c}.
		\end{cases}
\end{equation}
However, in the present fermionic model, it would be more accurate to replace the system size with the number of emitters, which is influenced by doping. Thus, we will introduce a rescaled $\alpha$ later on. Note that the calculations can be done before or after the rescaling. The only change to the Hamiltonian is a rescaling of the cavity energy and the trivial replacement $\alpha g \to \tilde{\alpha}\tilde{g}$ in the interaction term.

For simplicity, let us focus on the homogeneous case, where the electronic Hamiltonian is diagonalized by plane-wave modes with the dispersion relation
\begin{equation}
	E_{\nu,k} = -2J_\nu \cos(k) + \epsilon_\nu.
\end{equation}
In this case, the mean-field Hamiltonian splits into a direct sum of terms for each wavenumber, 
\begin{equation} \label{app:eq:H_MF_k}
	H(\alpha) = \omega_\text{c} \alpha^2 + \sum_k \left( E_{2,k}  b_{2,k}^\dagger b_{2,k} + E_{1,k} b_{1,k}^\dagger b_{1,k} + 2g\alpha \left[ b_{2,k}^\dagger b_{1,k} + b_{1,k}^\dagger b_{2,k} \right] \right).
\end{equation}
The eigenvalues are
\begin{equation}
	\begin{split}\label{app:eq:eigenvalues}
		E_k^\pm(\alpha) &= \overline{\epsilon}_k \pm K_k(\alpha) \\
		&= \frac{E_{2,k}+E_{1,k}}{2} \pm \sqrt{ \frac{ (E_{2,k}-E_{1,k})^2 }{4} + 4g^2\alpha^2 }
	\end{split}
\end{equation}
representing hybridized states of the upper and lower band. For convenience in the following, we also introduce
\begin{equation} \label{app:eq:abbreviations}
	\delta_k = \frac{E_{2,k} - E_{1,k}}{2}.
\end{equation}

To keep the derivation more general, and to express the partition function as a product of contributions from each wavenumber and orbital, we introduce a chemical potential $\mu$ and work in the grad canonical ensemble at finite temperature (the zero-temperature limit is derived later on). Thus, the partition function is given by
\begin{equation}
	\begin{split}
		Z(\alpha) 
		&= \Tr \left\{ e^{-\beta \left(H(\alpha) - \mu N\right)} \right\} \\
		&= e^{-\beta \omega_\text{c} \alpha^2 } \prod_{k} \left( 1 + e^{-\beta \left(E_{k}^+(\alpha) - \mu \right) } \right) \left( 1 + e^{-\beta \left(E_{k}^-(\alpha) - \mu \right) } \right).
	\end{split}
\end{equation}
We can then calculate the free energy $F(\alpha) = - \beta^{-1} \ln Z(\alpha)$ as
\begin{equation}
	\begin{split}
		F(\alpha) 
		&= \omega_\text{c} \alpha^2 - \frac{1}{\beta} \sum_{k} \ln
		\left( 
			1 + e^{-\beta \left( E_{k}^+(\alpha) + E_{k}^-(\alpha) - 2\mu \right) } 
			+   e^{-\beta \left(E_{k}^+(\alpha) - \mu \right) } + e^{-\beta \left(E_{k}^-(\alpha) - \mu \right) } 
		\right) \\
		&= \omega_\text{c} \alpha^2 - \frac{1}{\beta} \sum_{k} \ln \left( 
		2 e^{ -\beta \left( \overline{\epsilon}_k - \mu \right) }
		\left[
			\cosh\left( \beta \left( \overline{\epsilon}_k - \mu \right) \right) 
			+ \cosh\left( \beta K_k(\alpha) \right) 
		\right] \right),
	\end{split}
\end{equation}
where we have grouped the first two and last two terms from inside the bracket into a $\cosh$, respectively, and used the abbreviations from Eqs.~\ref{app:eq:eigenvalues}~and~\ref{app:eq:abbreviations}.

\subsection{Critical point of the superradiant transition}
The saddle-point condition for a critical point is 
\begin{equation}
	\left. \frac{\partial}{\partial\alpha} F(\alpha) \right|_{\alpha=0} = 0 \, , \qquad 
	\left. \frac{\partial^2}{\partial\alpha^2} F(\alpha) \right|_{\alpha=0} = 0.
\end{equation}
To evaluate these, one needs to compute the derivatives of $K_k(\alpha)$, i.e.
\begin{equation}\label{app:eq:derivatives}
		K_k'(\alpha) = \frac{4g^2\alpha}{K_k(\alpha)} \, , \qquad 
		K_k''(\alpha) = \frac{4g^2}{K_k(\alpha)} - \frac{4g^2\alpha K_k'(\alpha)}{K_k(\alpha)^2}.
\end{equation}
At $\alpha=0$, these simplify to
\begin{equation}
	K_k'(0) = 0 \, , \qquad 
	K_k''(0) = \frac{4g^2}{\delta_k}.
\end{equation}

For the first derivative of the free energy at $\alpha=0$, the situation is trivial, since $F'(0)=K_k'(0)=0$. On the other hand, for the second derivative, we obtain
\begin{equation}
	\begin{split}
		2\omega_\text{c} &= \sum_k \frac{K_k''(0) \sinh\left(\beta K_k(0)\right)}{\cosh\left( \beta \left( \overline{\epsilon}_k - \mu \right) \right) 
		+ \cosh\left( \beta K_k(0)\right)} \\
		&= 4g^2 \sum_k \frac{\sinh\left(\beta \delta_k \right)}{ \delta_k \left[ \cosh\left( \beta \left( \overline{\epsilon}_k - \mu \right) \right) 
		+ \cosh\left( \beta \delta_k \right) \right] },
	\end{split}
\end{equation}
which can be solved for the critical value $g_\text{c}$,
\begin{equation}
	g_\text{c} = \sqrt{ \frac{\omega_\text{c}}{2} \frac{1}{ \sum_k \frac{\sinh\left(\beta \delta_k \right)}{ \delta_k \left[ \cosh\left( \beta \left( \overline{\epsilon}_k - \mu \right) \right) 
	+ \cosh\left( \beta \delta_k \right) \right] } } }. \label{app:eq:gc_tight-binding}
\end{equation}

Let us compare this expression to the Dicke model for $N$ emitters,
\begin{equation}
	H^\text{Dicke} = \omega_\text{c} \alpha^2 + \sum_{j=1}^N \Delta \sigma_j^+ \sigma_j^- + 2g\alpha \sum_{j=1}^{N} \left( \sigma^+ + \sigma^- \right),
\end{equation}
which has its mean-field critical point at 
\begin{equation}
	g_\text{c}^\text{Dicke} = \sqrt{ \frac{\omega_\text{c}}{2 N} \frac{ \frac{\Delta}{2} \cosh\left(\frac{\beta\Delta}{2}\right)  }{ \sinh\left(\frac{\beta\Delta}{2}\right) } }.
\end{equation}
Note that $g_\text{c}\sim\sqrt{N}^{-1}$. Had the calculation been carried out using $\tilde{\alpha}$ and $\tilde{g}$ from Eq.~\ref{app:eq:rescaling}, the transition would be independent of the system size.

There are two essential differences between the two models. First, all two-level emitters of the Dicke model are degenerate with excitation energy $\Delta$, while the asymmetric dispersion of upper and lower band in the fermionic model lead to $k$-dependent excitation energies $2\delta_k$. Second, an additional term in the denominator of the denominator of Eq.~\ref{app:eq:gc_tight-binding} controls the total particle number through the chemical potential. 

In order to obtain a formula in terms of the total particle number instead of the chemical potential, we consider the zero-temperature limit in combination with a flat band limit. The flat band limit is justified, since the band gap is significantly larger than the bandwidth, but it has to be taken after the zero-temperature limit, because otherwise the bands would be either full or empty. In order to take the zero-temperature limit, we compare the arguments of the $\sinh$ and $\cosh$ in Eq.~\eqref{app:eq:gc_tight-binding}, since 
\begin{equation}\label{app:eq:zero_T_limit}
	\lim_{\beta\to\infty} \frac{\cosh\left(\beta(\overline{\epsilon}_k-\mu)\right)}{\sinh\left(\beta\delta_k\right)}
	= \begin{cases}
		\infty \qquad &\text{if } \left| \overline{\epsilon}_k-\mu \right| > \delta_k \\
		0 \qquad &\text{if } \left| \overline{\epsilon}_k-\mu \right| < \delta_k \\
		1 \qquad &\text{if } \left| \overline{\epsilon}_k-\mu \right| = \delta_k.
	\end{cases}
\end{equation}
Neglecting the last case of equality, the only contributions to the sum over $k$ in the denominator of Eq.~\eqref{app:eq:gc_tight-binding} come from $\left| \overline{\epsilon}_k-\mu \right| < \delta_k$, where 
\begin{equation}
	\lim_{\beta\to\infty} \frac{1}{\frac{\cosh\left(\beta(\overline{\epsilon}_k-\mu)\right)}{\sinh\left(\beta\delta_k\right)}+1} = 1.
\end{equation}
Correspondingly, the critical coupling of Eq.~\eqref{app:eq:gc_tight-binding} is simplified to
\begin{equation}\label{app:eq:gc_tight-binding_T0}
	g_\text{c} = \sqrt{ \frac{\omega_\text{c}}{2} \frac{1}{ 
		\sum_{\substack{k \text{ with}\\ \left| \overline{\epsilon}_k-\mu \right| < \delta_k}} 
		\frac{1}{ \delta_k } 
		} }. 
\end{equation}

The condition in the sum can be expanded as
\begin{equation}\label{app:eq:chem_pot_condition}
	\left| \overline{\epsilon}_k-\mu \right| < \delta_k  \Leftrightarrow  
	\begin{cases}
		E_{1,k} = \overline{\epsilon}_k - \delta_k < \mu \qquad \text{if } \overline{\epsilon}_k > \mu \\
		E_{2,k} = \overline{\epsilon}_k + \delta_k > \mu \qquad \text{if } \overline{\epsilon}_k < \mu.  
	\end{cases}
\end{equation}
We see that the chemical potential has to lie between the upper and lower Bloch state of a specific wave number, i.e., the lower Bloch state is occupied while the upper one is not. Therefore, there are $L(1-n_\text{d})$ terms in the sum, where $n_\text{d} = |N-L|/L$ is the doping density as defined in the main text. 

Finally, by taking the flat band limit $E_{\nu,k}=E_\nu$, which implies $\delta_k=\delta$, we arrive at the generalized scaling form of the critical coupling,
\begin{equation}\label{app:eq:gc_tight_binding_T0_flat_band}
	g_c \approx \sqrt{\frac{\omega_c \delta}{2} \cdot \frac{1}{L(1-n_\text{d})}} \quad \xrightarrow[\text{resonance}]{\; 2\delta = \omega_\text{c} \;} \quad \frac{\omega_\text{c}}{2} \frac{1}{\sqrt{L(1-n_\text{d})}}.
\end{equation}
By comparison, the simpler zero-temperature limit of the Dicke model is
\begin{equation}
	g_c^\text{Dicke} = \sqrt{\frac{\omega_c \Delta}{4 N}} \quad \xrightarrow[\text{resonance}]{\; \Delta = \omega_\text{c} \;} \quad \frac{\omega_\text{c}}{2} \frac{1}{\sqrt{N}}.
\end{equation}

With the rescaled coupling strength and order parameter from Eq.~\eqref{app:eq:rescaling}, one obtains critical point of the fermionic model that is independent of $L$, 
\begin{equation}
	\tilde{g}_c^\text{MF,SR} \approx \sqrt{\frac{\omega_c \delta}{2} \cdot \frac{1}{(1-n_\text{d})}}
\end{equation}

\subsection{Critical exponent $\beta$}
The correspondence to the Dicke model can be extended by showing that the superradiant transitions belong to the same universality class. Here, we compute just the critical exponent $\beta$, which relates the order parameter $\alpha$ to the coupling strength 
for $g>g_\text{c}$. By minimizing the free energy, $\frac{\partial}{\partial_\alpha} F(\alpha) = 0$, we obtain
\begin{equation}
	0 = \left( \frac{\omega_\text{c}}{2} \frac{1}{ \sum_k \frac{1}{K_k(\alpha)} \frac{\sinh\left(\beta K_k(\alpha)\right)}{  \left[ \cosh\left( \beta \left( \overline{\epsilon}_k - \mu \right) \right) 
	+ \cosh\left( \beta K_k(\alpha)\right) \right] } } - g^2  \right),
\end{equation}
where we have inserted the first derivative of $K_k(\alpha)$ from Eq.~\eqref{app:eq:derivatives}. 

Here, we can use the zero-temperature limit and flat band approximation as we did for the critical point. While the chemical potential depends on $\alpha$, the particle number is fixed, so we arrive at the same type of expression as in Eq.~\eqref{app:eq:gc_tight-binding_T0}, only with $K(\alpha)=K_k(\alpha)$ instead of $\delta$:
\begin{equation}\label{app:eq:alpha_g}
	\begin{split}
		0 &= \frac{\omega_\text{c}}{2} \frac{K(\alpha)}{ L(1-n_\text{d})} - g^2 \\
		&\approx g_\text{c}^2-g^2 + \frac{2\omega_c}{\delta L(1-n_\text{d})} \alpha^2 g^2.
	\end{split}
\end{equation}
In the second step, we have inserted the Taylor expansion of $K(\alpha)$,
\begin{equation}
	K_k(\alpha) \approx \delta_k \left( 1 + \frac{4g^2\alpha^2}{\delta_k^2} + \mathcal{O}\left(\left(\frac{4g\alpha}{\delta_k}\right)^4\right) \right)
\end{equation}
and $g_\text{c}^2=\omega_c \delta / (2L(1-n_\text{d}))$ from Eq.~\eqref{app:eq:gc_tight_binding_T0_flat_band}.

From Eq.~\eqref{app:eq:alpha_g} if follows that, in the vicinity of the critical point, $\alpha$ scales as
\begin{equation}
	\alpha \sim \sqrt{g^2-g_\text{c}^2} \sim \sqrt{g-g_\text{c}}.
\end{equation}
Therefore, we find a critical exponent $\beta=\frac{1}{2}$, just as in the Dicke model.

\end{document}